\documentclass{svjour3}                     
\smartqed  
\usepackage{graphicx}
\begin{document}

\title{SAI: a compact atom interferometer for future space
missions
}


\author{ F. Sorrentino \and
K. Bongs \and
P. Bouyer \and
L. Cacciapuoti \and
M. de Angelis \and
H. Dittus \and
W. Ertmer \and
A. Giorgini \and
J. Hartwig \and
M. Hauth \and
S. Herrmann \and
M. Inguscio \and
E. Kajari \and
T. K\"onemann \and
 C. L\"ammerzahl \and
 A. Landragin \and
 G. Modugno \and
 F. Pereira dos Santos \and
 A. Peters \and
 M. Prevedelli \and
 E. M. Rasel \and
 W. P. Schleich \and
 M. Schmidt \and
 A. Senger \and
 K. Sengstok \and
 G. Stern \and
 G. M. Tino \and
 R. Walser
}


\institute{G. M. Tino \and M. de Angelis \and F. Sorrentino \and A. Giorgini \at
             Dipartimento di Fisica, Universit\`a di Firenze, Polo Scientifico, via Sansone 1, I-50019 Sesto Fiorentino, Italy \\
              Tel.: +39 0554572034\\
              Fax: +39 0554572364\\
              \email{guglielmo.tino@fi.infn.it}           
	\and K. Bongs \at
	Midlands Ultracold Atom Research Centre
School of Physics \& Astronomy
University of Birmingham, Edgbaston, Birmingham B15 2TT, UK
	\and
	P. Bouyer \and G. Stern \at
	Laboratoire Charles Fabry de LÕInstitut dÕOptique, Centre National de la Recherche Scientifique, Campus Polytechnique Rd 128, 91127 Palaiseau, France
	\and
	L. Cacciapuoti \at
	European Space Agency, Research and Scientific Support Department, Keplerlaan 1, 2201 AZ Noordwijk, The Netherlands
	 \and H. Dittus \at
           Institute of Space Systems, German Aerospace Center (DLR), Robert-Hooke-Strasse 7, D Ð 28359 Bremen, Germany
           \and
          W. Ertmer \and E. M. Rasel \and J. Hartwig \at
          Institute of Quantum Optics, Leibniz Universit\"at Hannover, Welfengarten 1, D Ð 30167 Hannover, Germany
           \and
           M. Hauth \and A. Peters \and M. Schmidt \and A. Senger  \at
           Humboldt-Universit\"at zu Berlin, Unter den Linden 6, D-10099 Berlin, Germany
                     \and
         S. Herrmann \and T. K\"onemann \and C.   L\"ammerzahl \at
           University of Bremen, Centre of Applied Space Technology and Microgravity (ZARM), Am Fallturm, D - 29359 Bremen, Germany
           \and
           M. Inguscio \and G. Modugno \at
           European Laboratory For  Non Linear Spectroscopy (LENS), Via Nello Carrara, 1 50019 Sesto Fiorentino (FI), Italy
           \and
           E. Kajari 	\and W. P. Schleich  \at
	Institut f\"ur Quantenphysik, Universit\"at Ulm, Albert-Einstein-Allee 11, D-89081 Ulm, Germany
          \and
          A. Landragin \and F. Pereira dos Santos \at
          Observatoire de Paris - SYRTE 61 avenue de l'Obseravtoire, 75014 Paris, France
          \and
          M. Prevedelli \at
          Dipartimento di Fisica dell'Universit\`a di Bologna, Via Irnerio 46, I-40126, Bologna,  Italy
  \and
            K. Sengstok \at
            Universit\"at Hamburg, Edmund-Siemers-Allee 1, D-20146 Hamburg, Germany
            \and
            R. Walser \at   
	Institut f\"ur Angewandte Physik, Technische Universit\"at Darmstadt, Hochschulstr. 4a, D-64289 Darmstadt, Germany
              }

\date{}

\maketitle

\begin{abstract}
Atom interferometry represents a quantum leap in the technology for the ultra-precise monitoring of accelerations and rotations and, therefore, for all the science that relies on the latter quantities. These sensors evolved from a new kind of optics based on matter-waves rather than light-waves and might result in an advancement of the fundamental detection limits by several orders of magnitude.  Matter-wave optics is still a young, but rapidly progressing science.  The Space Atom Interferometer project (SAI), funded by the European Space Agency,  in a multi-pronged approach aims to investigate both experimentally and theoretically the various aspects of placing atom interferometers in space: the equipment needs, the realistically expected performance limits and potential scientific applications in a micro-gravity environment considering all aspects of quantum, relativistic and metrological sciences. A drop-tower compatible prototype of a single-axis atom interferometry accelerometer is under construction. At the same time the team is studying new schemes, e.g. based on degenerate quantum gases as source for the interferometer. A drop-tower compatible atom interferometry acceleration sensor prototype has been designed, and the manufacturing of its subsystems has been started. A compact modular laser system for cooling and trapping rubidium atoms has been assembled. A compact Raman laser module, featuring outstandingly low phase noise, has been realized.  Possible schemes to implement coherent atomic sources in the atom interferometer have been experimentally demonstrated.
\end{abstract}

\section{Introduction and motivation}
\label{intro}
Matter-wave interferometry has recently led to the
development of new techniques for the measurement of inertial
forces, with important applications both in fundamental
physics and applied research. The remarkable stability
and accuracy that atom interferometers have reached for
acceleration measurements can play a crucial role for
science and technology. Quantum sensors based on atom interferometry
had a rapid development during the last decade and different measurement
schemes were demonstrated and implemented. Atom
interferometry is used for precise measurements of the gravitational
acceleration \cite{Kasevich92,Peters99,Muller07}, EarthÕs gravity gradient \cite{Snadden98,McGuirk02}, and rotations
\cite{Gustavson97,Gustavson00}. Experiments on  the validity of the equivalence principle \cite{Fray04} and on the
measurement of the gravitational constant G \cite{Bertoldi06,Fixler07,Tino08} are currently in progress, while
 tests of general relativity \cite{Dimopoulos06} and of
Newton's $1/r^2$ law \cite{Tino03,Ferrari06} as well as the detection of gravitational waves
\cite{Tino04,Dimopoulos08,Dimopoulos09} have been proposed.
Accelerometers based on atom interferometry have
been developed for many practical applications including
metrology, geodesy, geophysics, engineering prospecting and
inertial navigation \cite{McGuirk02,Peters01,Leone06}. Ongoing studies show that the
space environment will allow us to take full advantage of the
potential sensitivity of atom interferometers \cite{Tino07,Turyshev07}.

Space Atom Interferometer (SAI) is a project of the European Space Agengy (ESA contract n. 20578/07/NL/VJ,
AO-2004-064/082), with the main goal to demonstrate the possibility of applying such technology to future space missions.

\subsection{Cold atom sensors}
\label{atominterferometer}

In analogy to optical interferometers, atomic matter-waves
can be split and recombined giving rise to an interference signal. Different schemes have been used for splitting,
reflecting and recombining atomic matter waves.
In the SAI project, the Raman interferometer is chosen as the basic configuration \cite{Kasevich92}. 
This approach was proven to be one of the most promising concepts in atom interferometry, at least for classical atom sources with respect to applications in precision experiments. 

\begin{figure}
  \includegraphics[width=0.95\textwidth]{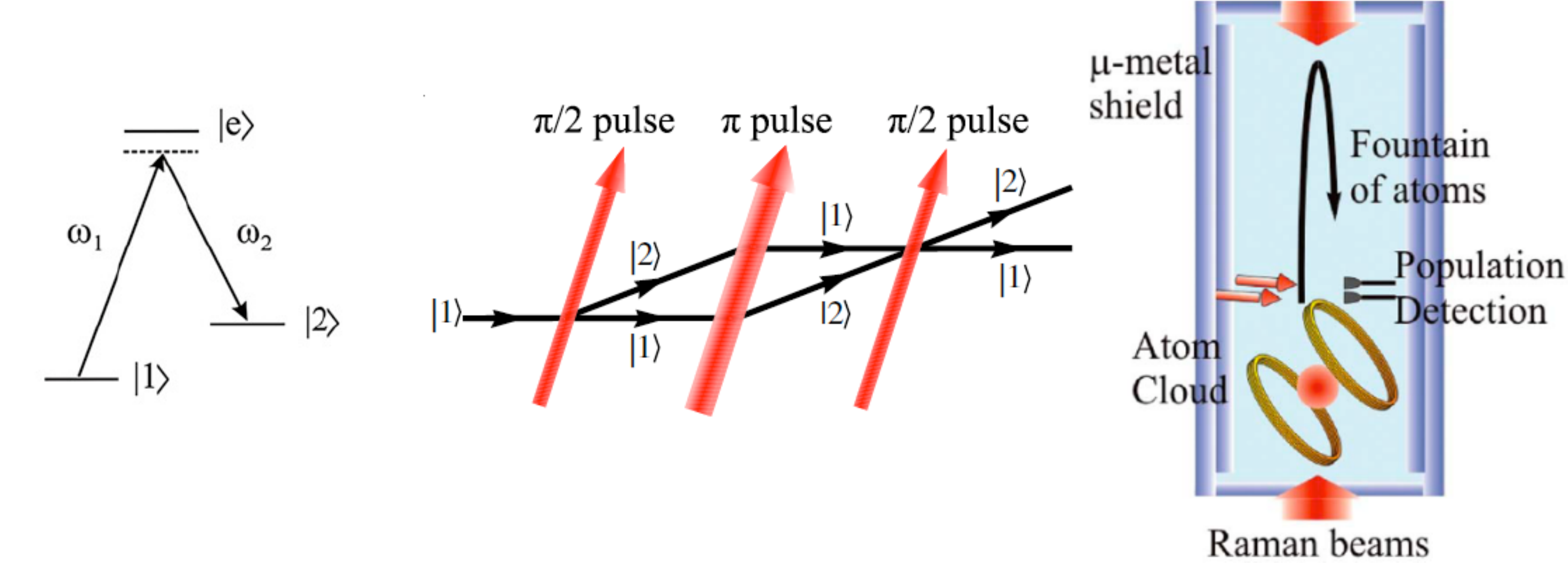}
\caption{Atom interferometry accelerometer; from left to right: Raman transition between the hyperfine levels of an alkali atom; Mach-Zender atom interferometry scheme implemented with three Raman pulses ($\pi/2-\pi-\pi/2$); typical scheme of an atom interferometry sensor operated under the Earth gravity: an atomic cloud is first trapped and cooled in an ultra-vacuum chamber, then is launched in free flight and interacts with a Raman beams pair, and finally interacts with resonant probe beams for fluorescence population detection.}
\label{Ramanconcept}       
\end{figure}

Fig. \ref{Ramanconcept} illustrates the basic principles of a matter-wave interferometry sensor based on laser cooled alkali atoms. Raman transitions change the internal hyperfine state and at the same time the external state of the atom by 2 photon recoils. A sequence of three atom-light interaction zones can be applied to coherently split, redirect and recombine the atomic de Broglie wave. After the first beam splitter two atomic de Broglie waves emerge. Their relative velocities differ by the photon recoil transferred to the atom. Choosing the duration of the atom-light interaction properly one may (as in the first and the third zone) equally split the incident waves (so called $\pi/2$ pulse), or (as in the second zone) deflect entirely the incident waves ($\pi$ pulse). The two exit ports of the interferometer could be addressed and detected separately by spectroscopy, like the excitation by laser and detection of the fluorescence.

The enormous sensitivity of atomic sensors depends on the measurement time $T$, during which the matter wave accumulates the phase along the interferometer trajectories (see section \ref{baseline}). Laser cooling is therefore required to reach lower temperatures for a maximum extension of the measurement time $T$ as well as a high efficiency to achieve large numbers, $N$, of cold atoms. In absence of gravity the duration of the interaction is limited by the drift time of the atoms through the experimental apparatus, which could be in principle tens of seconds for the lowest atomic temperatures (nK) presently reached. On Earth, gravity is the major problem for cold atom optics: the free fall of about 5\,m in one second restricts the measurement time well below 1\,s. 
%


\subsection{The transportable atom interferometer}
\label{transportable}
The main goal of the SAI project is to demonstrate the technology readiness of atom interferometry inertial sensors for space applications. For this purpose, a compact prototype of single-axis accelerometer based on ultracold $^{87}$Rb has been designed and is currently being assembled. At the same time, the project is investigating new schemes, based for example on quantum degenerate gases as source for the interferometer. 

The realization of portable devices will be the first step towards the development of space instruments based on matter wave interferometry. In addition, it will enable comparisons between atomic inertial sensors at a level not achievable with other commercial devices - due to their very high level of sensitivity and accuracy (at the limit of the quantum noise). 
Finally, SAI investigates emerging concepts such as pulsed and continuous atom lasers. With the most sensible devices at hand, the project provides a unique framework to test the classical concepts of interferometry in combination with these non-classical sources and to develop new strategies for employing this new form of quantum matter.
Design target is to keep the prototype as compact as possible without degradation of sensitivity as compared to existing laboratory instruments. 
Besides portability, several properties of the sensor envisage future implementation in microgravity. The prototype has been be designed to be a vertical accelerometer with the capability of launching the atoms vertically, in a fountain-like configuration. Goal sensitivity to accelerations of  $\sim1\times10^{-7}$\,m/s$^2$ ($3\times10^{-7}$\,m/s$^2$ specified) at 1\,s of integration time will be reached during ground testing of the instrument. Operating the same instrument in microgravity would allow a $\sim100$-fold improvement in the acceleration sensitivity, using an interaction time $T$ of the order of 1\,s.

\section{Baseline structure of the SAI sensor}
\label{baseline}
Space experiments impose strict limitations on the volume of the apparatus, the weight, and its power consumption. The final selection of materials, mechanical design, and implementation strategy takes into account all these requirements. 
The SAI sensor is a compact 1-axis accelerometer with sensitivity in the range of $10^{-8}$\,g at 1\,s. 
A modular laser system will be connected through optical fibers to a sensor head consisting of an ultra-high vacuum system with attached optics and of a magnetic shield.
The same vacuum chamber is used for both trapping and cooling the atoms and to detect them at the end of the interferometric sequence. The possibility of launching the atomic sample is important for ground testing of the instrument. This scheme is mandatory to demonstrate possible future operation in microgravity. The implementation of a 2D-MOT \cite{Dieckmann98} as source of cold atoms for a standard 3D-MOT ensures ultra-high vacuum conditions in the loading and interferometer regions, important for operating the sensor with coherent atomic sources (BEC) and for testing its performance at long interrogation times.


The output phase of a Raman pulse matter-wave accelerometer is \cite{Peters99}
\begin{equation}
\phi=kaT^2
\end{equation}
where $k$ is the wavevector associated to the Raman transitions ($1.6\times10^7$\,m$^{-1}$ in Rb), $a$ is the acceleration and $T$ is the time interval between Raman pulses. The phase resolution $\Delta\phi$ depends on the signal-to-noise ratio at detection; for high enough number of atoms, detection is limited by quantum projection noise and the phase resolution per shot is $\delta\phi\approx1/\sqrt{N}$, thus the acceleration resolution per shot is
\begin{equation}
\delta a=\frac{1}{\sqrt{N}kT^2}.
\end{equation}

Typical measurement cycle with $\sim5$\,Hz repetition rate would be
\begin{itemize}
\item launch $\sim10^8$ atoms after loading for $\sim100$\,ms from the 2D-MOT;
\item velocity and internal state selection resulting in $\sim10^6$ atoms;
\item interferometer sequence with duration $2T\sim100$\,ms;
\item detection of $\sim10^5$ atoms with SNR $\sim100$ (3 times worse than QPN);
\end{itemize}
resulting in virtual acceleration sensitivity of $10^{-7}$\,m/s$^2$ at 1\,s.
In order to demonstrate technology readiness for space applications, the sensor is designed to fit into the drop-tower capsule of the Bremen drop tower. The mechanical structure holding the vacuum system and all necessary optics shall be contained into a cylindrical volume with diameter $\sim60$\,cm and height $\sim150$\,cm.

\section{Sensor subsystems}
\label{subsystems}

Main subsystems of the SAI sensor are:
\begin{itemize}
\item mechanical structure of the sensor;
\item vacuum system with a single chamber for atom trapping and interrogation/detection;
\item high-flux source of atoms (2D-MOT);
\item laser source system for atom trapping and manipulation;
\item Raman laser system for the atom interferometer operation;
\item electronic control system.
\end{itemize}

\subsection{Mechanical structure and main vacuum system}
\label{mechanical}

\begin{figure*}
  \includegraphics[width=0.75\textwidth]{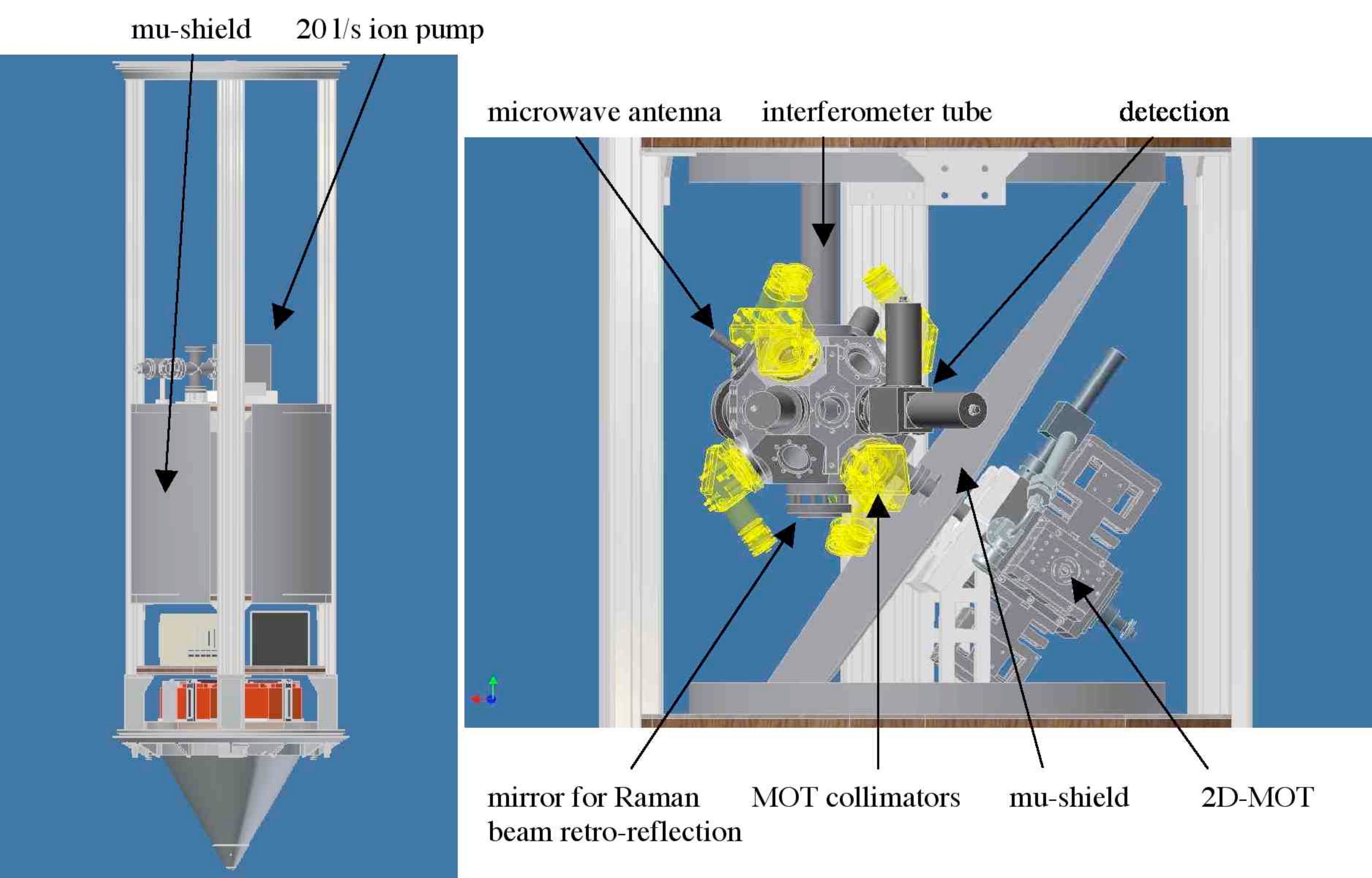}
\caption{3D views of the SAI sensor fitting into a drop-tower capsule}
\label{capsule}     
\end{figure*}

The SAI sensor will be a single-axis accelerometer. We have technically designed the mechanical structure of the SAI sensor, including magnetic shields. Main goal of the design work was to make the whole SAI system fit into the ZARM standard drop capsule (long version).
Fig. \ref{capsule} shows a 3D view of  the SAI sensor integrated in the drop-tower capsule. The main cell with a 20\,cm tube for the interferometer sequence, the 2D-MOT system and the fiber collimators are well visible. 
The maximum launch height of 300\,mm corresponds to a free flight time lower than 500\,ms. 
Launch geometry in 1-1-1 configuration is chosen, with a short vertical tube above main chamber, which allows to launch the atoms along the vertical and to operate the sensor as a gradiometer for performance measurements. Since the interferometer sequence has to happen close or within the main chamber, magnetic shielding of the tube would not be sufficient. A $\mu$-metal shield will enclose the whole vacuum system. 

\subsubsection{Main vacuum system}
\label{vacuum}

\begin{figure*}
  \includegraphics[width=0.75\textwidth]{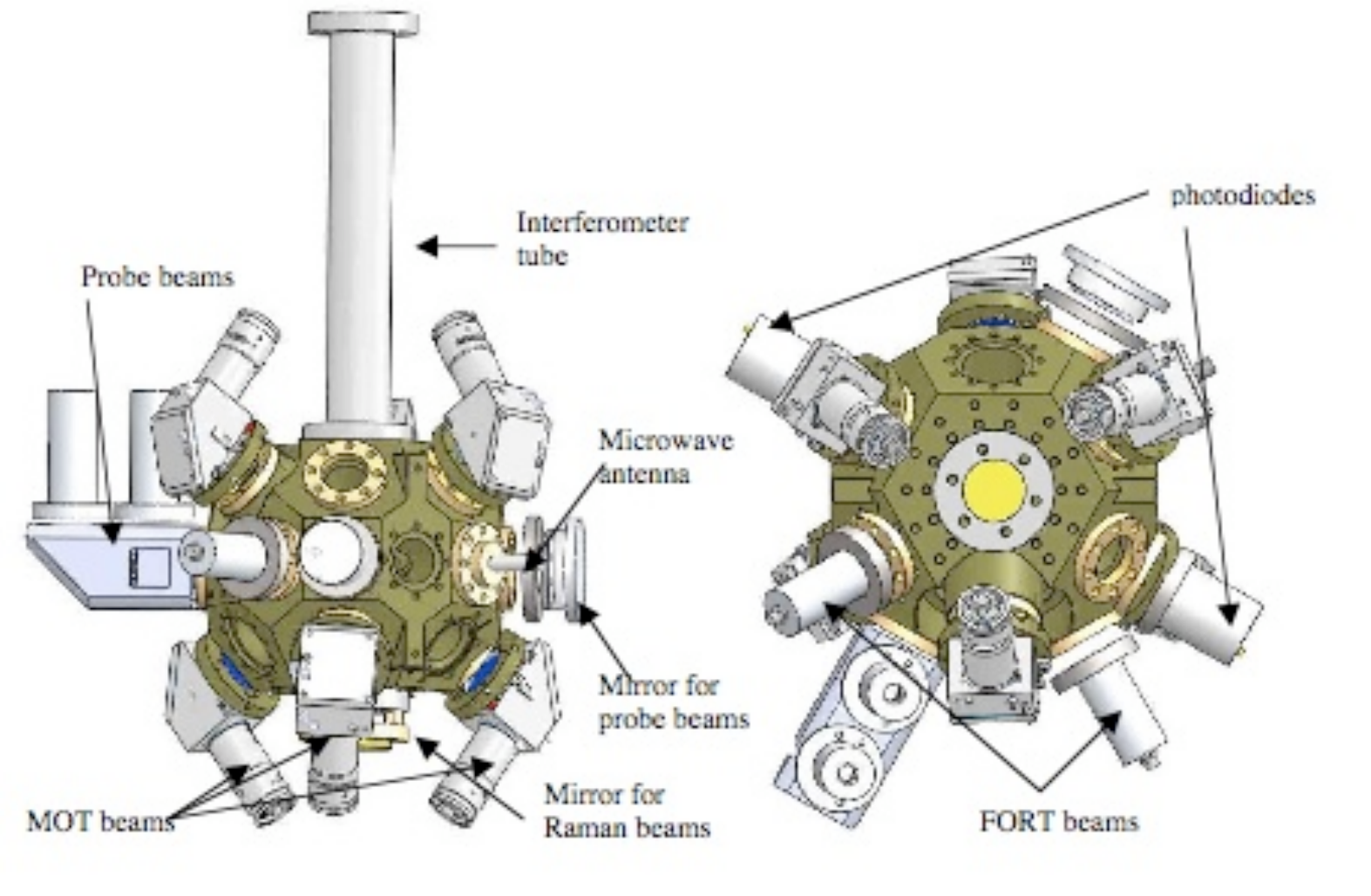}
\caption{3D views of the main vacuum chamber}
\label{chamber}     
\end{figure*}

The main vacuum chamber is shown in fig. \ref{chamber}. It has 24 view-ports giving access for cooling laser beams, 2D-MOT atomic flux, laser beams for optical BEC, detection optics. In order to comply with geometrical constraints given by a drop tower capsule, folded collimators are used for laser beam delivery from fibres to the cell. Raman beams will propagate into the same polarization maintaining optical fibre, and will be delivered to the main chamber from a collimator placed over the upper tube; the retro-reflecting mirror will be attached below the main chamber. A precise tiltmeter will be rigidly attached to the mirror. Microwave excitation of atoms will be induced by an antenna placed outside a 30 mm diameter window in the horizontal symmetry plane of the main chamber. An ion pump for the main vacuum system will be attached at the upper end of the 20 cm tube.
Optical BK7 windows will be attached with indium wire sealing. This technology has already proved to be compatible with the acceleration loads of the Bremen drop tower \cite{Quantus}.

The vacuum system is contained in the central part of the drop-tower capsule, between two horizontal platform that are rigidly connected to four vertical stringers. All kind of implemented components are rigidly mounted on the capsule platforms to withstand decelerations up to 50\,g during the impact of the drop capsule at the end of the free fall. The main vacuum chamber is rigidly connected to the upper capsule platform through the vertical vacuum tube, that is made in titanium and has a wall thickness of 2.5 mm. The tube is clamped to the platform. Between the mechanical contact of the main chamber and 2D-MOT vacuum cell an oscillation compensator (bellow) is inserted to additionally protect this ultra-high vacuum connection from longitudinal impact forces. The 2D-MOT system is rigidly connected to the lower capsule platform through an aluminum support attached to the massive square vacuum connector.

\subsubsection{Magnetic shield}
\label{shield}


Magnetic field gradients can introduce systematic shifts in the interferometer phase readout. Moreover, fluctuations in the magnetic field direction would influence the launching direction of atoms, degrading the sensor stability. Typical peak-to-peak magnetic field variations of 500 mG can be expected along the ~100 m tube of the Bremen drop tower. These fields are strongly attenuated by the SAI  $\mu$-metal shield assembly. The shield consists in a set of two concentric boxes of high magnetic permeability material. An attenuation factor of 60\,dB  (both radially and axially) in the central part of the tube, where the interferometer sequence will take place, is expected. The static field generated by the ion pump attached at the end of the 20\,cm titanium tube is attenuated by an additional 2\,mm tick  $\mu$-metal shield. The main vacuum camber and the 2D-MOT assembly are separated by a double layer  $\mu$-metal wall, in order to get rid of the effect of metallic parts in the commercial Rb dispenser and to attenuate the static field generated by the small (2\,l/s) ion pump attached to the 2D-MOT. The remaining metallic parts of 2D-MOT system are made of aluminum. All mechanical components inside the main magnetic shield (main chamber, vacuum tubes, optical mounts) are carefully chosen to be non-magnetic. 
Care has been taken to limit magnetic effects in the interferometer region, using the following considerations:
\begin{itemize}
\item strong permanent magnets (in ion pumps) are separately shielded;
\item all mechanical components within main shield are non-magnetic;
\item holes in main shield are small compared to their distance from the interferometer region.
\end{itemize}

\begin{figure*}
  \includegraphics[width=0.75\textwidth]{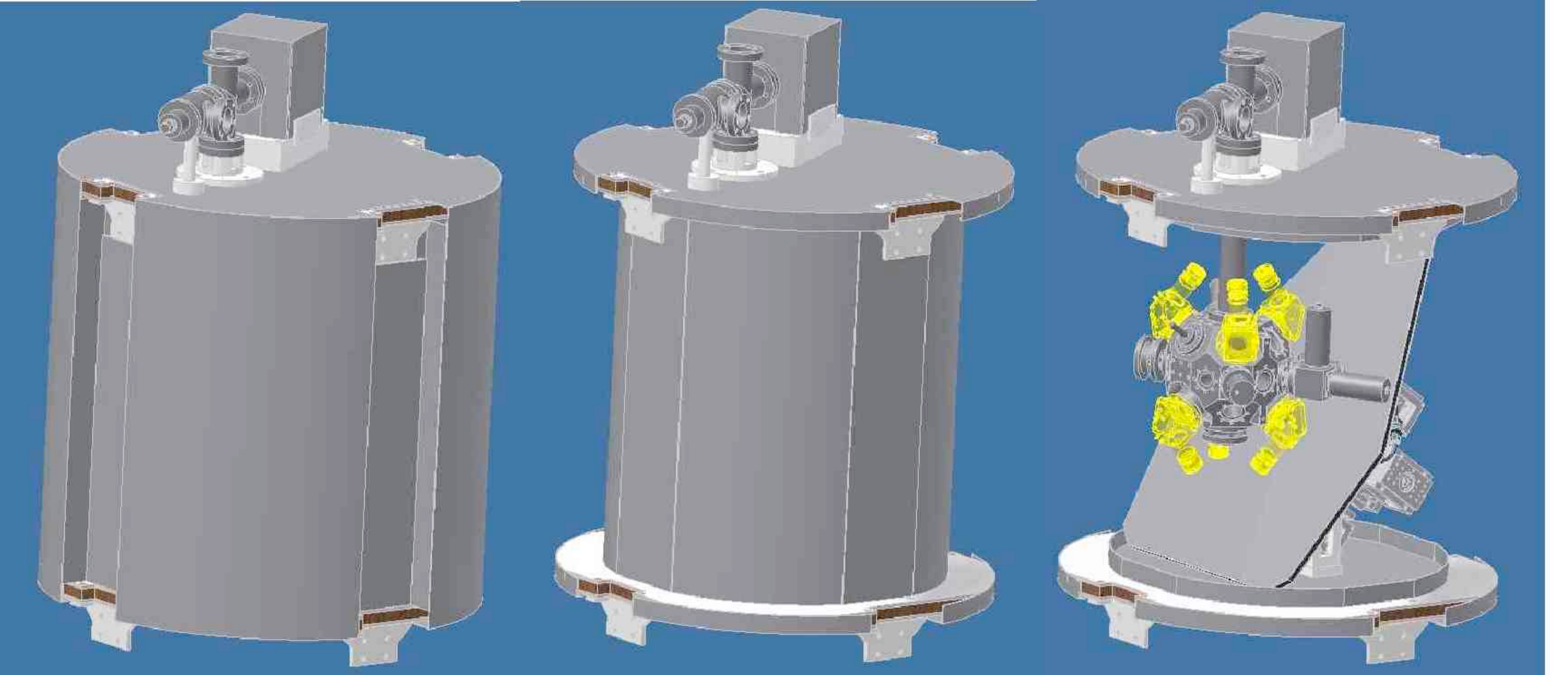}
\caption{3D views of the magnetic shield; from left to right: outer view; with outer cylindrical
layer removed; with inner cylindrical layer removed.}
\label{shield3D}     
\end{figure*}

As a consequence of the geometrical constraints given by the drop-tower capsule, the shape of the magnetic shield is not symmetric. Analytical formulas are not suited to precisely describe the magnetic field distribution within the shielded region. In order to find the optimal shape for the magnetic shield, we implemented a numerical model based on a commercial software (ANSYS). Goal of our analysis was to keep the residual field, in the region where the atom interferometry takes place, below 300 $\mu$G with an external applied field of 0.5 G in either direction. As a result, the shape shown in fig. \ref{shield3D} is well suited for our purposes. The shield consists of two nearly cylindrical $\mu$-metal layers with 2\,mm thickness, enclosing the whole vacuum system, and a double diagonal plane to separate the 2D-MOT from the main chamber. The distance between the two layers is 50 mm. 

%

\begin{figure*}
  \includegraphics[width=0.5\textwidth]{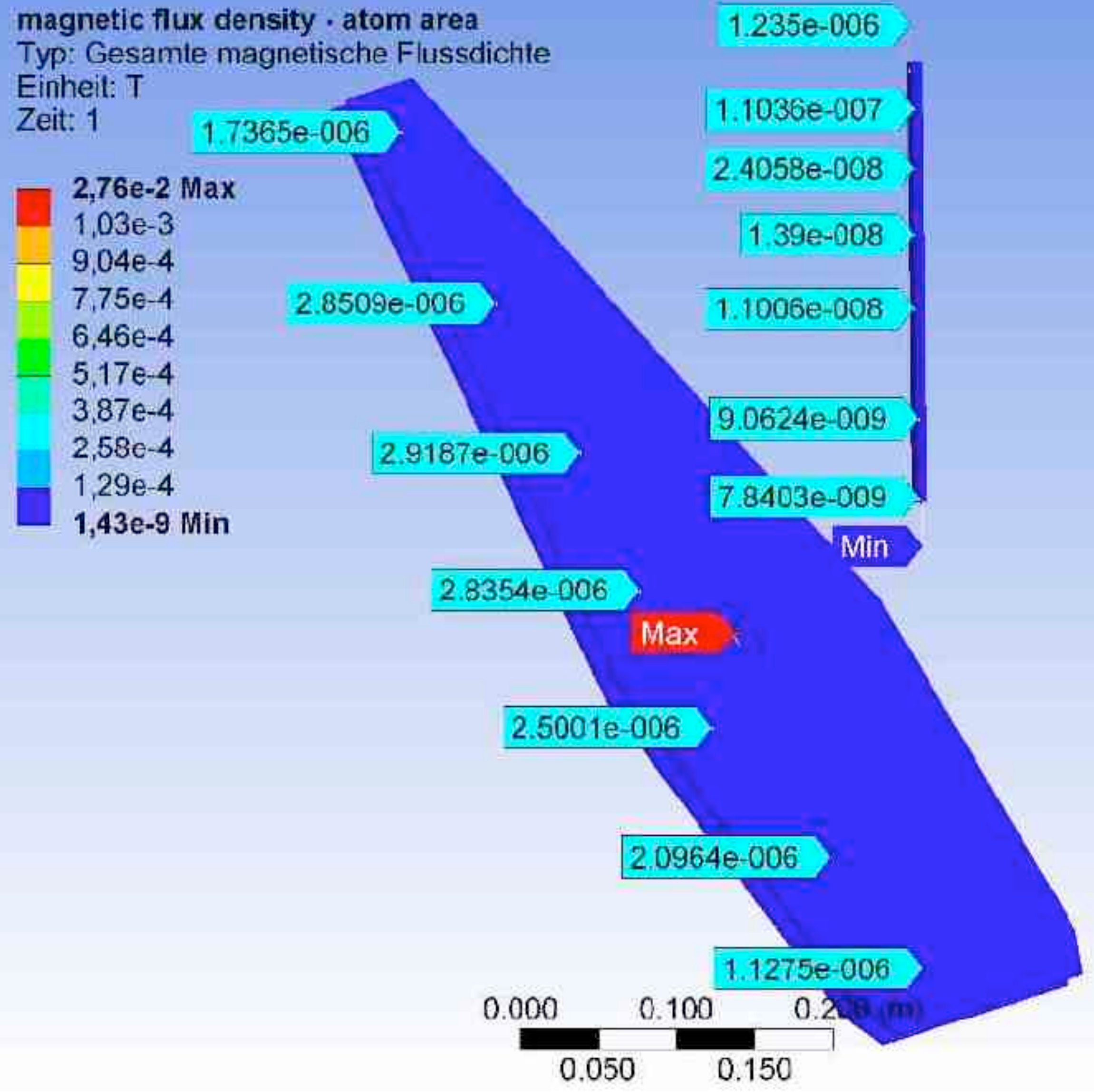}
\caption{Calculated residual magnetic field along the atomic trajectory when an external bias field of 10\,G is applied along the vertical axis.}
\label{ANSYS}     
\end{figure*}

The mechanical structure, including magnetic shield, must be qualified for drop tower experiments. This gives additional constraints to the shape of magnetic shield. The mounting of $\mu$-metal layers must be rigid enough to withstand 50 g accelerations. Moreover, the shield cannot enclose the four vertical capsule stringers. The shape of magnetic shield 
has been chosen to fulfill such requirements by keeping the distance between $\mu$-metal layers large enough to produce the desired shielding factor.
The top and bottom covers  of the cylindrical $\mu$-metal layers are simply attached to the sides of the horizontal platforms, while the cylindrical layers and the diagonal walls are clamped to the platforms and to the stringers in several points.
Typical results of our numerical calculations are illustrated in fig. \ref{ANSYS}, which shows the computed field intensity distribution (expressed in T) for an external applied field of 500 mG in the vertical direction.

According to our calculations, the magnetic field would be uniform in time and in space within 300\,$\mu$G along the atom interferometry region during a free fall in the drop tower. The calculated shielding factor at the center of the main vacuum chamber has a minimum value of 6000 along the vertical direction, due to the large aperture on the Raman beams path. The same analysis has been used to find a configuration of coils to produce a uniform bias field for the atom interferometry sequence.

The overall mass of the magnetic shield will be about 85 Kg.

\subsection{High flux atomic source}
\label{2DMOT}

The 2-DMOT for the Space atom interferometer will be a mixture of the design implemented in the CASI experiment \cite{CASI} for cold-atom Sagnac interferometer and the ATLAS experiment \cite{ATLAS} for an all optical produced Bose-Einstein condensate.

\begin{figure*}
  \includegraphics[width=1.0\textwidth]{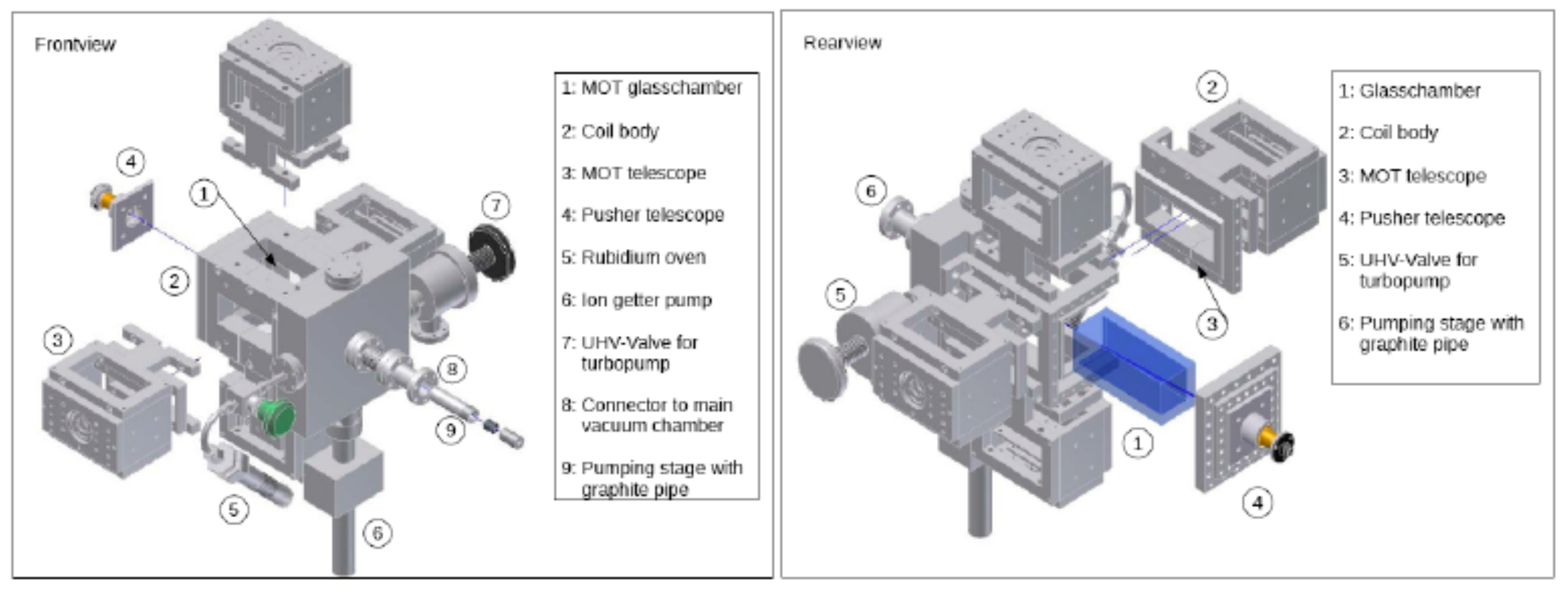}
\caption{3D views of the mechanical structure of the 2D-MOT system.}
\label{2DMOT3D}     
\end{figure*}

The design of the SAI 2D-MOT system mechanical structure is based on a glass cell pressed on a non-magnetic aluminium body (see fig. \ref{2DMOT3D}). The aluminium body will provide the connections for 
\begin{itemize}
\item ion pump with a pumping speed of 2\,l/s
\item connector for a Turbomolecular pump (pre-vacuum pump, only used during assembly)
\item Rb reservoir with a valve (CASI layout)
\item belllow to the actual interferometer chamber. 
\end{itemize}

The 2D-MOT of ATLAS and CASI are not equipped with a pump. CASI offers the possibility of initiation by a turbomolecular pump and uses a heatable Rb reservoir. ATLAS uses dispensers rather than Rb reservoirs. Both CASI and ATLAS use a rigid connector to the main chamber. Differently from the above mentioned experiments, the SAI 2D-MOT will be connected to the main chamber via a bellow and it will have an additional 2\,l/s ion pump.
The 2D-MOT structure also serves as mount for the 4 telescopes, shaping the laser beams for cooling the atoms in the transverse direction to the atomic beam, and for the magnetic coils.

\subsection{Laser system}
\label{lasersystem}

\begin{figure*}
  \includegraphics[width=0.95\textwidth]{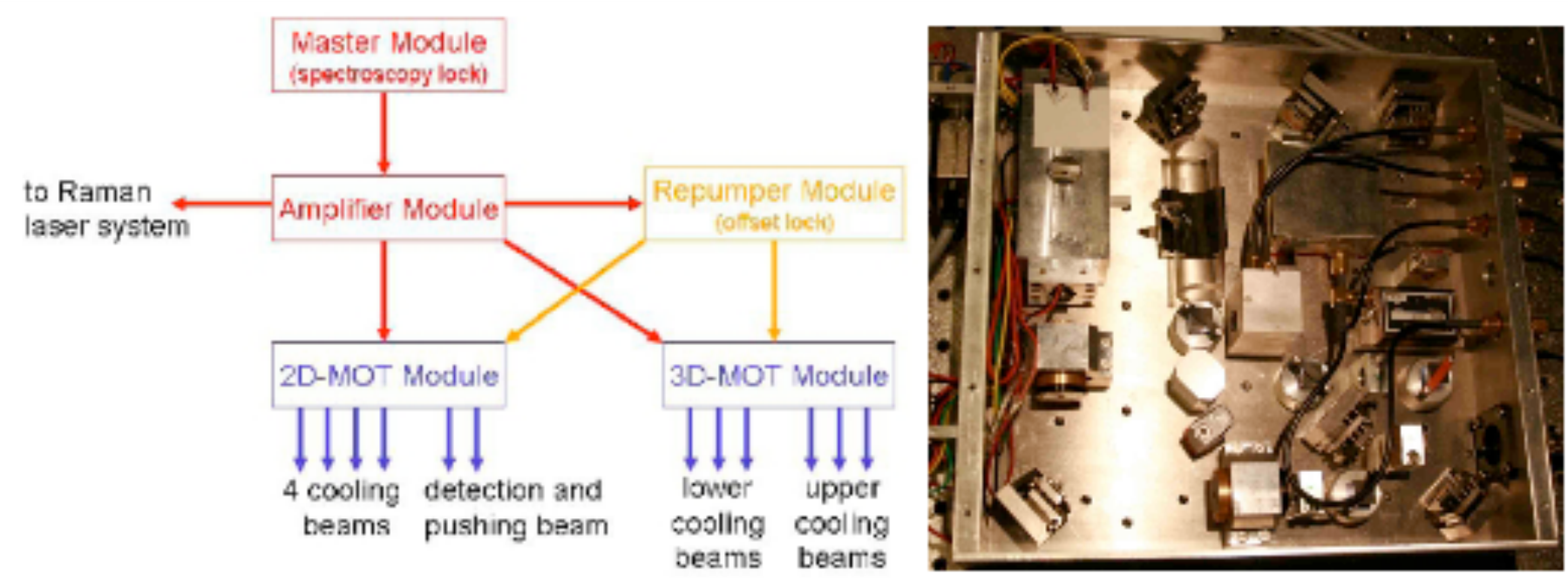}
\caption{Left: scheme of the modular laser system; right: a picture of the master laser module.}
\label{modules}     
\end{figure*}

The laser system for cooling and detecting Rb atoms is subdivided into five compact modules shown in fig. \ref{modules}. Raman laser beams for the interferometer sequence are produced in an additional module, described in section \ref{Ramanlasersystem}.

In the master module, a laser is stabilised on the Rb FM spectroscopy signal and sent to a second module, where it is amplified. This light is used as a reference for an offset-lock in the repumper module and in the Raman laser system; in addition, it is also forwarded to the 2D-MOT and 3D-MOT laser modules. The two MOT modules shift the master frequency to the cooling frequency and amplify it, in addition allowing a control of the laser detuning from resonance and of laser power. Finally, they distribute the cooling and repumping light onto several outputs. The arrows represent optical fibre connections.

Each module was designed and optimised for compactness and robustness. The beam height was set to only 2\,cm and mounts and frames created of robust aluminium. The mounts had already been developed and tested  under drop-tower conditions in the framework of the QUANTUS project \cite{Quantus}.
Fig. \ref{modules} also shows the 
completed frame with its optical components of the master module. The Laser (ECL) is frequency-shifted via an AOM and afterwards stabilized by Doppler-free observation of a cross-over resonance in the $^{87}$Rb glass cell with a photodiode.

The Amplifier, Repumper, the two MOT modules, as well as the Raman laser module described in section \ref{Ramanlasersystem}, were designed, manufactured and assembled using similar components, mounts and technology. The dimensions of all modules are listed in table \ref{lasermodules}.

Extended Cavity Lasers (ECLs) are used in  the SAI laser system. They employ interference filters for wavelength selection. These lasers, based on an interferential filter for wavelength selection \cite{Baillard06},
offer the narrow linewidth of conventional grating-based ECLs, but they are more than one order of magnitude less sensitive to misalignment. This characteristic makes them ideal for implementation in mobile laser systems, where mechanical stability is of paramount importance.


\begin{table}
\caption{Laser module dimensions 
}
\label{lasermodules}       
\begin{tabular}{llll}
\hline\noalign{\smallskip}
Module & Width [mm] & Depth [mm] & Height [mm]  \\
\noalign{\smallskip}\hline\noalign{\smallskip}
Master Module & 236 & 250  & 63 \\
Amplifier Module & 240 & 190  & 66 \\
Repumper Module & 186 & 260  & 63 \\
2D-MOT Module & 350 & 260  & 73 \\
3D-MOT Module & 350 & 260  & 73 \\
Raman Module & 430 & 430 & 106 \\
\noalign{\smallskip}\hline
\end{tabular}
\end{table}

%

%

%


%

\subsection{Raman laser system}
\label{Ramanlasersystem}

For inducing a two-photon optical Raman transition on the hyperfine ground state of rubidium, a pair of two lasers with a fixed phase relation, a frequency difference of approximately 6.8 GHz and a wavelength of 780 nm is required. The most obvious method to achieve this is to phase lock two lasers onto each other. An optical phase-locked loop (OPLL) has become the standard for atom interferometry experiments in various laboratories and is also the baseline setup for SAI.
%

\subsubsection{Optomechanical implementation and OPLL}
\label{optomechanical}


%

The Raman laser optical setup 
contains two ECLs. The Raman master laser is overlapped on a photodiode with the reference laser generated by the module already described in sec. \ref{lasersystem}. The resulting beat note signal is locked on a frequency reference provided  by the frequency synthesis chain described in sec. \ref{chain}. The second ECL (slave) is overlapped with the master laser and their beat note is phase locked to the low phase noise signal at 6.8\,GHz (hyperfine splitting of $^{87}$Rb) provided by a frequency synthesis chain (see section \ref{chain}). Since the phase difference between Raman master and slave lasers is imprinted directly on the atoms, phase noise performance of both the frequency reference and the optical phase locked loop are of key importance for the sensor performance. It is therefore imperative to minimize the residual phase noise between the two lasers. The specified acceleration sensitivity of SAI ($3\times10^{-7}$\,m/s$^2$ in 1\,s) corresponds to a maximum allowed phase noise spectral density of  -85\,dBrad$^2$/Hz between 10\,Hz and 10\,kHz.
Both the master and the slave laser beams are amplified in two separate tapered amplifiers (TA).   

As shown in fig. \ref{chain_spectrum}, the phase noise spectral density for the OPLL stays below a level of -120\,dBrad$^2$/Hz (1 $\mu$rad/$\sqrt{\hbox{Hz}}$) between 100\,Hz and 60\,kHz, well compatible with the noise level required by SAI. To our knowledge, the measured OPLL phase noise is among the lowest levels ever reached. The achieved locking bandwidth is slightly above 4\,MHz. 
If this OPLL phase noise was the main contributing factor to overall accelerometer sensitivity, a single-shot sensitivity $\delta a/a$ of $2\times10^{-10}$ would be achieved.


\subsubsection{Reference frequency chain}
\label{chain}

A microwave frequency source is used to generate the reference for the optical phase-locked loop (OPLL) of the Raman lasers. With typical bandwidths of the OPLL of several MHz, the phase stability of the microwave source reduces drastically the relative phase fluctuations of the lasers. Still, small phase fluctuations of this reference source will limit the sensitivity of the interferometer, in a way that depends on the interferometer parameters (interferometer duration, duration of the Raman pulses, etc.). As the repetition rate of an atom interferometer experiment lies between a fraction of a Hz and few Hz, requirements on the phase noise at low frequency are stringent, and state of the art oscillators are necessary. 
A conventional solution consists in generating a microwave reference by multiplication of ultra low noise quartz oscillators. This signal can then be compared directly to the optical beat note, or can be used to phase lock a microwave oscillator such as a YIG or a DRO onto the quartz signal. This last solution offers the possibility to realize a tunable frequency source, whose phase and/or frequency can be easily modulated. Also, it improves the phase noise at high frequencies.
Better performances can be obtained using sapphire oscillators. 

The contribution of the microwave source onto the phase noise of the interferometer can be reduced down to about 1 mrad/shot for $2T=100$ ms, using commercial ultra low noise quartz oscillators of various classes (BlueTop from Wenzel, or BVA from Oscilloquartz). When extrapolating the performances of an atom interferometer to larger interrogation times ($2T\leq1$\,s), the phase noise at low frequency will be the dominant contribution. The BVA quartz appears here as the best solution among the different types of commercial low phase noise quartz oscillators. Assuming a phase noise scaling as $1/f^2$ for frequencies below 10\,Hz, the expected sensitivity per shot of the interferometer scales linearly with $T$. We calculate from the specifications of the OCXO 8607 from Oscilloquartz an interferometer phase noise of 9\,mrad/shot for $T=1$\,s.  As for the microwave frequency synthesis, its contribution can be reduced to less than 1\,mrad/shot by a careful design, independent of $T$.

The source consists in a reference signal at 100\,MHz, obtained by phase locking a 100\,MHz onto a BVA 5\,MHz quartz oscillator. This double quartz system is necessary to obtain the lowest level of phase noise both in the low and high frequency range. 

The 100\,MHz signal is then be multiplied by 2, amplified and sent to a Non Linear Transmission Line (NLTN), which generates a comb of frequencies, harmonics of 200\,MHz. The 6.8\,GHz tooth of the comb is  then filtered, amplified, and used to phase lock a Dielectric Resonator Oscillator (DRO) with an offset frequency provided by a Direct Digital Synthesizer (DDS1). The DRO is thus phase and frequency tunable. A second DDS is used as the reference of the OPLL.

The multiplication stage is therefore composed of two stages: 
\begin{itemize}
\item a Radio Frequency (RF) stage with a $\times2$ multiplier (RK3+ from MiniCircuits), a band pass filter at 200\,MHz (BL Microwave B0200-100C1BS), an RF amplifier (ERA 5 from MiniCircuits), a low pass passive filter at 250\.MHz (MiniCircuits);
\item a microwave (MW) stage 
with a comb generator (NLTL from Picosecond Pulse Labs), a band pass filter at 6.8\,GHz (BL Microwave BP6800-70-6-CS or 7\,GHz BP7000-70-6-CS), a MW amplifier (AML612L3401 from AML Communications).
\end{itemize}
The band pass filter in the MW stage is narrow enough to select only one line of the frequency comb (the bandwidth of the filter is 70\,MHz).

\begin{figure*}
  \includegraphics[width=0.95\textwidth]{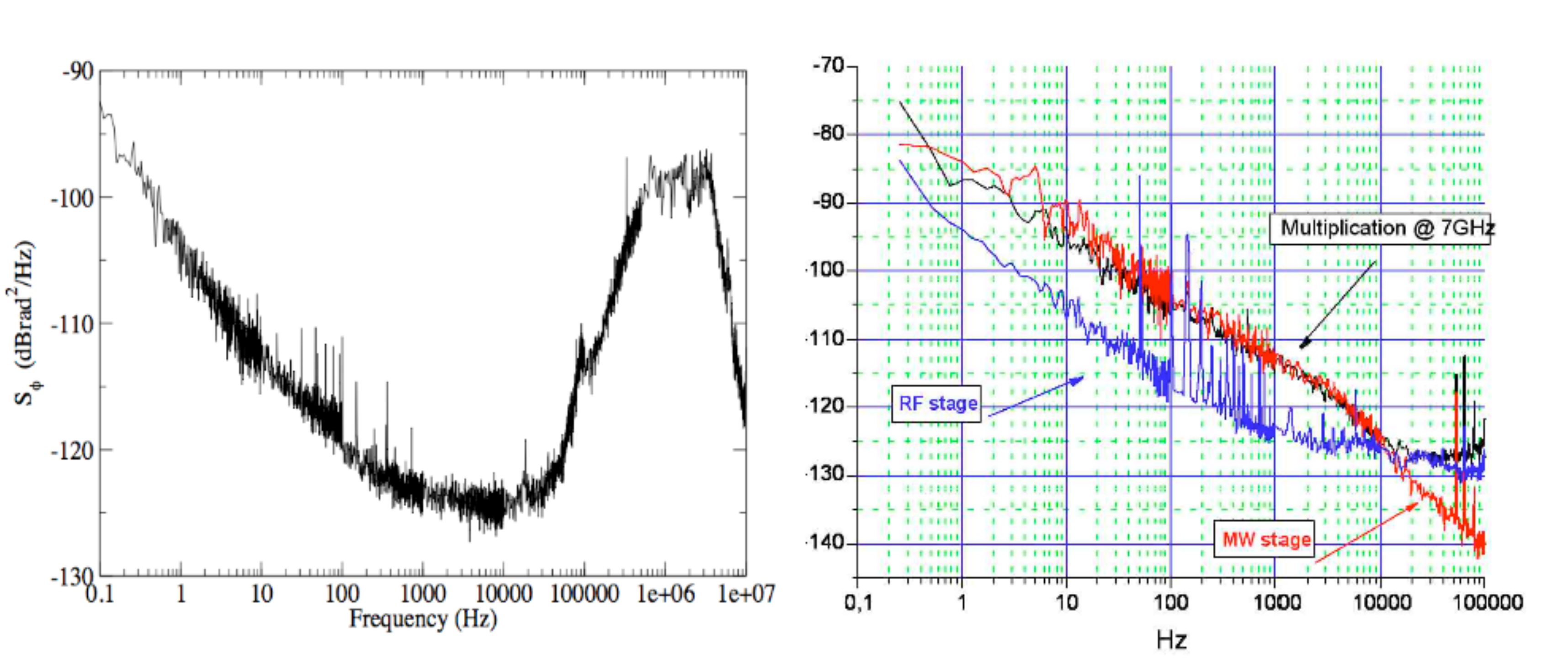}
\caption{Left: Raman laser OPLL phase noise; right: measured phase noise of the 6.8 GHz frequency chain.}
\label{chain_spectrum}     
\end{figure*}

For a frequency offset from the carrier lower than 10\,kHz, the RF stage limits the global phase noise of the multiplication stage. For a larger frequency  the noise introduced by the MW stage appears (cf. fig. \ref{chain_spectrum}).

%

\subsection{Electronic control system}
\label{controlsystem}

We have developed a versatile source of signals able to control the different phases for the production and the analysis of a cold atom experiment. It is based upon a series of parallel modules, generating parallel signals with only one clock. In this case, each output of each module is generated in a dedicated component: a Field Programmable Gate Array (FPGA).
This device can be integrated in a stand-alone 19'' rack.
It uses an Interface Control Module (ICM) to load programming data and controls from  the serial port interface of a PC. It can accept up to 8 Signals Generator Modules (SGM). Depending on its type, each module can supply up to 16 output channels. Each channel can output a sequential 32-step signal. Each step can be programmed from 1\,$\mu$s to 10\,s duration without any restriction on the other channels.
The overall control system of the SAI system will consist in three analog modules (12 analog output
channels), three TTL modules (24 TTL output channels) and 8 single-channel DDS
boards.

\subsection{Coherent atomic source}
\label{coherent}

 Baseline design of the SAI sensor includes integration of a coherent atomic source.
The coherent source under development in SAI will produce a BEC of $^{87}$Rb atoms with an all-optical approach using a fiber laser at 1565\,nm \cite{Clement09}. All-optical techniques allow fast productions of BEC compared with magnetic traps, and optical potentials can trap atoms whatever their spin. The chosen method is finally quite simple: neither additional cooling (such as Raman sideband cooling), nor magnetic fields are required, nor movable optical parts to compress the optical trap. Only molasses and CMOT phases are needed to get a good loading before performing an evaporation stage until the phase transition. The keypoint of such setup is the specific light-shift features at this wavelength for $^{87}$Rb. Transitions from $^5P_{3/2}$ to $4D$ states at 1529\,nm induce strong light-shift for the $^5P_{3/2}$ levels. Light-shift of the upper level of the cooling transition  $^5P_{3/2}$ is nearly 42 times stronger than the light-shift of the ground state. The effect of 1565\,nm light on the cooling transition allows a loading procedure with simultaneous cooling and trapping.

%

\section{Conclusions}
\label{conclusions}

We have designed a compact and transportable prototype of atom interferometry accelerometer, to demonstrate the possibility to apply such technology in future space missions. The system in under construction, and several subsystems have already been completed and tested. The Space Atom Interferometer project will provide a valuable insight into the potential of atom interferometry for space applications.

\begin{acknowledgements}
This work was supported by the European Space Agency through the contract n. 20578/07/NL/VJ. A. G. and F. S. acknowledge financial support by the European Science Foundation through the Euroqasar program.
\end{acknowledgements}



\end{document}